# Outstanding Thermal Conductivity of Single Atomic Layer Isotope-Modified Boron Nitride

Qiran Cai,[1] Declan Scullion,[2] Wei Gan,[1] Alexey Falin,[1] Pavel Cizek,[1] Song Liu,[3] James H. Edgar,[3] Rong Liu,[4] Bruce C.C. Cowie,[5] Elton J. G. Santos,[6] Lu Hua Li[1]*

1. Institute for Frontier Materials, Deakin University, Waurn Ponds Campus, Waurn Ponds, VIC 3216, Australia.

2. School of Mathematics and Physics, Queen's University Belfast, Belfast BT7 1NN, United Kingdom.

3. Tim Taylor Department of Chemical Engineering, Kansas State University, Manhattan, Kansas 66506, USA.

4. Advanced Materials Characterisation Facility, University of Western Sydney, Penrith, NSW 2751, Australia.

5. Australian Synchrotron, 800 Blackburn Road, Clayton, VIC 3168, Australia.

6. School of Physics and Astronomy, The University of Edinburgh, EH9 3FD, United Kingdom.

*Email: luhua.li@deakin.edu.au

ABSTRACT

Materials with high thermal conductivities ($\kappa$) is valuable to solve the challenge of waste heat dissipation in highly integrated and miniaturized modern devices. Herein, we report the first synthesis of atomically thin isotopically pure hexagonal boron nitride (BN) and its one of the highest $\kappa$ among all semiconductors and electric insulators. Single atomic layer (1L) BN enriched with $^{11}$B has a $\kappa$ up to 1009 W/mK at room temperature. We find that the isotope engineering mainly suppresses the out-of-plane optical (ZO) phonon scatterings in BN, which subsequently reduces acoustic-optical scatterings between ZO and transverse acoustic (TA) and longitudinal acoustic (LA) phonons. On the other hand, reducing the thickness to single atomic layer diminishes the interlayer interactions and hence Umklapp scatterings of the out-of-plane acoustic (ZA) phonons, though this thickness-induced $\kappa$ enhancement is not as dramatic as that in naturally occurring BN. With many of its unique properties, atomically thin monoisotopic BN is promising on heat management in van der Waals (vdW) devices and future flexible electronics. The isotope engineering of atomically thin BN may also open up other appealing applications and opportunities in 2D materials yet to be explored.

A dramatic temperature rise can occur locally at the so-called "hot spot" in highly integrated and miniaturized devices, including microprocessor and circuit packages, light-emitting diodes, high-power lasers and radio frequency transmitters. Eliminating waste heat limits the performance, reliability, and longevity of many modern devices. Materials with outstanding thermal conductivities ($\kappa$) is one of the key solutions to this challenge. While diamond and graphite are the only two traditional materials with $\kappa$ above 1000 W/mK at room temperature, emerging materials such as carbon nanotubes [1], graphene [2], and recently discovered cubic boron arsenide (*c*BAs) (~1000 W/mK) [3-5] are also excellent thermal conductors. However, electrically conductive carbon materials are not suitable in direct contact with electronic devices due to the potential for short circuiting. Although diamond and *c*BAs have the potential for high-power electronics, they are unsuitable for flexible electronic devices and new two-dimensional (2D) van der Waals (vdW) structures.

Distinct from their carbon counterparts, all boron nitride (BN) materials, including single-wall nanotubes and single atomic layer or monolayer (1L) BN are electric insulators and hence better candidates for waste heat dissipation, for example, in electronic devices. Bulk cubic (*c*) and hexagonal (*h*) BN crystals are good thermal conductors with $\kappa$ of ~690 and 420 W/mK at room temperature, respectively [6]. Recently, we reported that high-quality and surface-clean 1L *h*BN had a $\kappa$ of 751±340 W/mK [7]. This $\kappa$ increase with reduced thickness down to the atomic level was due to a decrease in the number of phonon branches and states available for Umklapp scattering with less interlayer interaction. Defects, grain boundaries, and surface contaminations, nevertheless, could adversely affect the thermal conduction of atomically thin BN [8-12].

Isotope engineering affects many fundamental properties of a solid, *e.g.* lattice parameter, disordering, elastic constant, vibration, band structure and transition, exciton, polariton dispersion and scattering. It, in turn, gives rise to appealing phenomena and applications, including the elevation of superconducting transition temperature, improvement in the lifetime of organic light-emitting diodes (OLED), optical fibers with higher speed, precise and accurate quantification of proteomes, and ultra-trace environmental analysis [13-15]. Naturally occurring BN ($^{Nat}$BN) contains a relatively high percentage of two stable boron isotopes: 19.9% $^{10}$B and 80.1% $^{11}$B; while carbon (C) normally consists of 98.9% $^{12}$C and only 1.1% $^{13}$C. The phonon energy, electronic bandgap, and electron density distribution of *h*BN could be varied by isotope engineering [16]. Isotope enriched *h*BN greatly increased polariton lifetime [17]. In addition, $^{10}$B is one of the best neutron absorbers and used widely in radiation shielding, nuclear reactivity control, and neutron capture therapy for tumor treatment [18,19]. Replacing $^{10}$B by $^{11}$B, on the other hand, prevents electronic devices from data loss or single-event upset caused by cosmic rays or their generation of ionizing particles.

Reducing isotopic disorder also increases thermal conductivity. The in-plane $\kappa$ of isotopically pure $^{12}$C graphene is 36% and ~100% higher than that of naturally occurring graphene and graphite, respectively [20]. The $\kappa$ of 1L isotopically pure $^{100}$MoS$_2$ were 61.6 W/mK, larger than the 40.8 W/mK of 1L $^{Nat}$MoS$_2$ [21]. Note that chemical vapor deposition (CVD) was used to synthesize these monoisotopic graphene and MoS$_2$. In terms of BN, the room temperature $\kappa$ of bulk $^{10}$BN crystals was 585 W/mK, ~39% higher than that of bulk $^{Nat}$BN [22]. The effect of isotopic impurity on the $\kappa$ of BN nanotubes was also studied: 310 W/mK for $^{11}$BN nanotubes, much larger than the 200 W/mK of $^{Nat}$BN nanotubes as a control [23]. Very recently, an ultrahigh $\kappa$ of 1600 W/mK was achieved from isotopically enriched *c*BN, ~190% higher than that of natural occurring *c*BN [24]. However, there has been no report on the synthesis of atomically thin monoisotopic BN, let alone measurement of its $\kappa$, though a 25-36% enhancement in $\kappa$ was theoretically predicted from 1L isotopically pure BN compared to that of 1L $^{Nat}$BN [25-28].

In this work, we successfully produced atomically thin isotopically pure $^{10}$BN and $^{11}$BN for the first time, and their intrinsic in-plane thermal conductivities could be determined due to their high quality and clean surface. Based on optothermal Raman measurements, the $\kappa$ of 1L $^{11}$BN and $^{10}$BN were 1009±313 and 958±355 W/mK, respectively. These values were ~34% and ~140% larger than those of 1L and bulk $^{Nat}$BN, respectively. Density functional theory (DFT) simulations were used to gain insights into the isotope effect. This study may also give rise to new possibilities in many other applications, *e.g.* multifunctional metal-matrix nanocomposites for radiation shielding and new cancer treatment [18,19].

High-quality and surface-clean atomically thin isotopically pure BN sheets were mechanically exfoliated from bulk crystals grown by the nickel-chromium solvent method [16,29]. According to secondary ion mass spectrometry (SIMS), these bulk crystals contained 99.2% and 99.9% $^{10}$B and $^{11}$B, respectively, close to the previously reported values [16]. Naturally occurring nitrogen has >99.6% $^{14}$N, and thus can be considered as isotopically pure. The atomically thin $^{10}$BN and $^{11}$BN sheets were directly exfoliated and suspended over pre-fabricated micro-wells (3.8 μm in diameter) connected by narrow trenches (200 nm in width) in 80 nm gold-coated Si substrates (Au/Si) without polymer-assisted transfer process. The Au films served as heat sinks during measurements; the trench acted as vents to avoid strain induced to the atomically thin materials due to volume change of the air trapped in the micro-wells during heating.

FIG. 1a and b show the atomic force microscopy (AFM) image and height trace of a suspended 1L $^{10}$BN with a thickness of 0.50 nm. More samples and AFM images are shown in Supporting

Information (FIG. S1). The chemical composition, crystal structure, and quality of the isotopically pure samples were probed by near-edge X-ray absorption fine structure (NEXAFS) spectroscopy, and compared with those of a single crystal $^{Nat}$BN synthesized by the high pressure Ba-BN solvent method (FIG. 1c) [30]. Sharp π* resonances at 192.0 eV corresponding to 1s core electron transitions to the unoccupied antibonding orbitals of B atoms $sp^2$-bond to three nitrogen atoms were observed from all samples, verifying their hexagonal crystal structure. No satellite peaks caused by other chemical environments were present, suggesting high chemical purities of the isotopically pure samples [31,32]. These results are in line with the previous finding that the $^{10}$BN and $^{11}$BN crystals were free of defects in the areas of tens of microns [22].

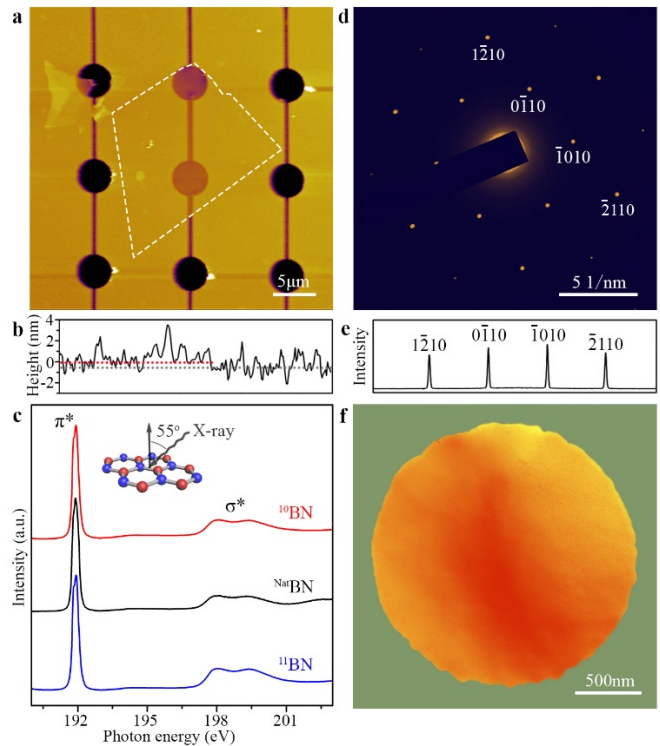

FIG. 1. (a) AFM image of a mechanically exfoliated 1L $^{10}$BN suspended over Au/Si substrate with micro-wells and trenches; (b) AFM height of the 1L sample; (c) NEXAFS spectra in the B K-edge region of isotopically pure and natural BN at the incidence of 55°; (d) selected area electron

diffraction of a suspended 1L $^{10}$BN; (e) the corresponding diffraction spot intensities; (f) the DF-TEM image.

Transmission electron microscope (TEM) was used to detect whether the exfoliated atomically thin sheets contained any grain boundary. FIG. 1d shows a typical electron diffraction pattern from the entire suspended area of a 1L $^{10}$BN on perforated silicon nitride (SiN$_x$) TEM grid with 2 μm holes. The diffraction pattern consisted of single sets of diffraction dots with six-fold symmetry. The intensity profile of the $(1\bar{2}10)$, $(0\bar{1}10)$, $(\bar{1}010)$ and $(\bar{2}110)$ diffraction reflections is plotted in FIG. 1e, which can be compared with those of 2L and few-layer samples shown in Supporting Information, FIG. S2. The dark field (DF)-TEM image also revealed a mono-crystalline nature without grain boundaries (FIG. 1f). The TEM results of a 1L $^{11}$BN are available in Supporting Information (FIG. S3).

FIG. 2a compares the Raman spectra of suspended 1-3L and bulk $^{10}$BN, $^{Nat}$BN and $^{11}$BN. Since the Raman frequency is inversely proportional to the square root of mean atomic mass, the bulk $^{10}$BN, $^{Nat}$BN, and $^{11}$BN crystals showed G bands centered at 1392.0, 1366.9, and 1358.0 cm$^{-1}$, respectively. Reducing the thickness of suspended monoisotopic BN to the atomic scale barely changed their G band Raman frequencies but lowered the peak intensities. A similar phenomenon on atomically thin $^{Nat}$BN was reported and explained by us before [33,34]. Due to mass disorder effects, the different isotope mass also affected the full width at half maximum (FWHM) of the G bands of bulk $^{10}$BN, $^{Nat}$BN, and $^{11}$BN crystals, *i.e.* 5.9, 9.4, and 5.6 cm$^{-1}$, respectively [35]. The atomically thin sheets showed broader bandwidths, caused by stronger surface scattering influencing the vibrational excitation lifetime [33,34].

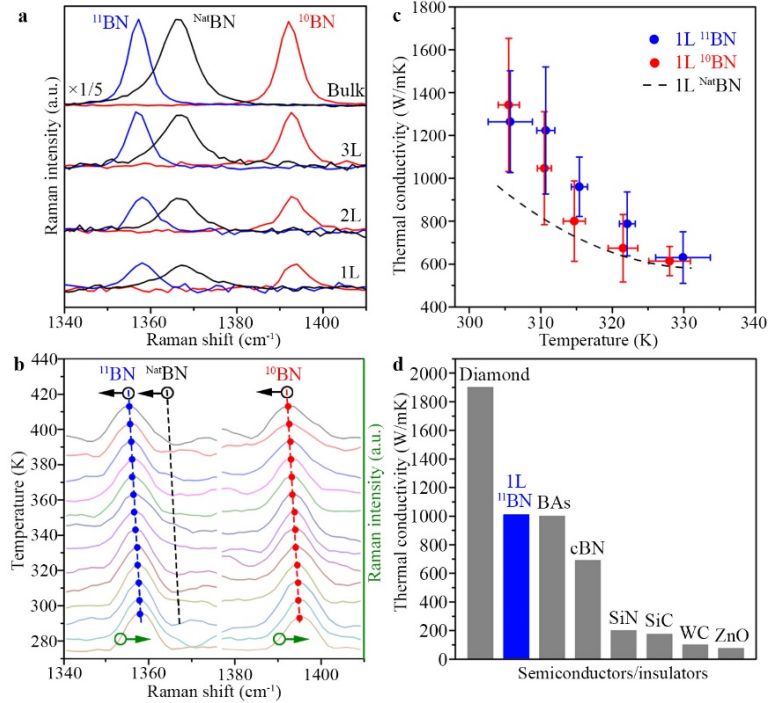

FIG. 2. (a) Comparison among Raman spectra of 1-3L and bulk $^{10}$BN, $^{Nat}$BN, and $^{11}$BN; (b) temperature effect on the Raman G band frequency of 1L $^{10}$BN, $^{Nat}$BN and $^{11}$BN sheets with corresponding fittings; (c) experimental $\kappa$ of 1L $^{10}$BN (red) and $^{11}$BN (blue) with standard deviations as a function of temperature, compared with that of 1L $^{Nat}$BN (black and dashed); (d) the comparison among $\kappa$ of representative semiconductors and insulators.

The $\kappa$ of atomically thin $^{10}$BN and $^{11}$BN was measured by optothermal Raman technique [2,7,21,36-40]. First, the Raman G band frequency of the suspended 1L $^{10}$BN and $^{11}$BN as a function of temperature was determined using a hot plate with accurate temperature control (±0.1 °C). In order to minimize laser heating, a small laser power of ~1.5 mW was chosen. FIG. 2b summarizes the Raman G bands of a 1L $^{10}$BN and $^{11}$BN at 293-413 K with an interval of 10 K and the corresponding linear fittings, i.e. $\omega - \omega_0 = \chi T$, where $\chi$ is the first-order temperature coefficient, and $\omega - \omega_0$ is the change of the G band frequency due to temperature variation. The $\chi$ values of 1L $^{10}$BN and $^{11}$BN were −0.0223±0.0018 (red dashed line in FIG. 2b) and −0.0220±0.0022 (blue dashed line) cm$^{-1}$/K, respectively, quite similar to that of $^{Nat}$BN, i.e. −0.0223±0.0012 cm$^{-1}$/K (black dashed line) [7]. Note that the volumetric thermal expansion of the Au/Si substrate during heating hardly affected these $\chi$ values due to the hanging down of suspended atomically thin BN, releasing the strain induced by thermal expansion coefficients mismatch between BN and the substrate, as we described before [7].

The suspended 1L $^{10}$BN and $^{11}$BN sheets were then optically heated under different laser power (4-10 mW) to increase the local temperature ($T_m$) (see Supporting Information, FIG. S4 for the Raman spectra). Given that the heat flux vector is along the radial direction away from the center of the suspended BN sheets and the phonon transport is diffusive, the thermal conductivity was calculated by:

$$\kappa = \frac{\ln(\frac{R}{r_0})}{2\pi t \frac{T_m - T_a}{Q - Q_{air}}} \alpha \quad (1)$$

where $R$ is the radius of the micro-well (1.9 μm); $r_0$ is the radius of the laser beam which was 0.32±0.01 μm (see Supporting Information, FIG. S5); Here, the laser beam radius ($r_0$) was the position where the laser intensity decreases to 1/e of its maximum intensity. The definition of $r_0$ at either 1/e or 1/e$^2$ intensity does not affect the obtained value of thermal conductivity (see Supporting Information). $\alpha = 0.97$ is the Gaussian profile factor of the laser beam; $t$ is BN thickness; $T_m$ is the temperature measured by Raman; $T_a$ is the ambient temperature; $Q$ is the absorbed laser power [20,37]. The optical absorption of 1L $^{10}$BN and $^{11}$BN at 488 nm wavelength was determined by the difference in the measured laser power between empty and nearby BN-covered holes of SiN$_x$ grids (see Supporting Information, FIG. S6). That is, Q = P$_{empty}$−P$_{BN}$. There was no noticeable difference in the absorbance of 1L $^{10}$BN and $^{11}$BN, and the averaged value was (0.32±0.13)%, close to that of 1L $^{Nat}$BN [7]. $Q_{air}$ is the heat loss in the air:

$$Q_{air} = \int_{r_0}^{R} 2\pi h(T - T_a)r\,dr + \pi r_0^2 h(T_m - T_a) \quad (2)$$

where $T$ is the temperature at radius $r$; $h$ is the heat transfer coefficient of $h$BN. In the case of small temperature variation between an object and the ambient, the quadratic expression for radiation can be simplified to the linearized sum of convective ($h_c$) and radiative ($h_r$) components to obtain the total heat transfer coefficient. That is, $h = h_c + h_r$, where $h_c$=3475W/m$^2$K for BN sheets; $h_r = \varepsilon\sigma 4T^3$; $\varepsilon = 0.8$ is the emissivity of $h$BN; and $\sigma$ is the Stefan-Boltzmann constant with the value of 5.670373×10$^{-8}$ W/m$^2$K$^4$ [41].

The $\kappa$ of 1L $^{10}$BN and $^{11}$BN as a function of temperature was calculated based on Equation 1, and compared with that of 1L $^{Nat}$BN from our previous study (FIG. 2c) [7]. The errors were calculated through the root sum square error propagation approach, where the temperature calibration by Raman, temperature resolution of the Raman measurements, and the uncertainty of the measured laser absorbance were considered. Due to the small temperature range and the uncertainty of the optothermal technique, we averaged the $\kappa$ values: 958±355 and 1009±313 W/mK for 1L $^{10}$BN and $^{11}$BN near room temperature, respectively. The heat loss to air only accounted for about 1.8% of the total heat dissipation during laser heating, indicating the use of the emissivity of $h$BN did not affect the calculated $\kappa$ of atomically thin BN much. Note optical heating was coupled more strongly to diffusive phonons of higher frequency than ballistic phonons, and the temperature measured by the Raman method was the anharmonic scattering temperature between the zone-center or zone-boundary optical phonons and diffusive acoustic phonons [38,42,43]. In addition, the local non-equilibrium of phonon polarizations was ignored [44]. As a result, these Raman-deduced $\kappa$ values should be underestimated. Our results showed that the $\kappa$ of 1L monoisotopic BN was about 34% and 140% higher than those of 1L $^{Nat}$BN and bulk $^{Nat}$BN, respectively [6,7], though the $^{Nat}$BN had a slightly lower destiny of defects. FIG. 2d compares the $\kappa$ of 1L $^{11}$BN with those of some representative semiconductors and insulators.

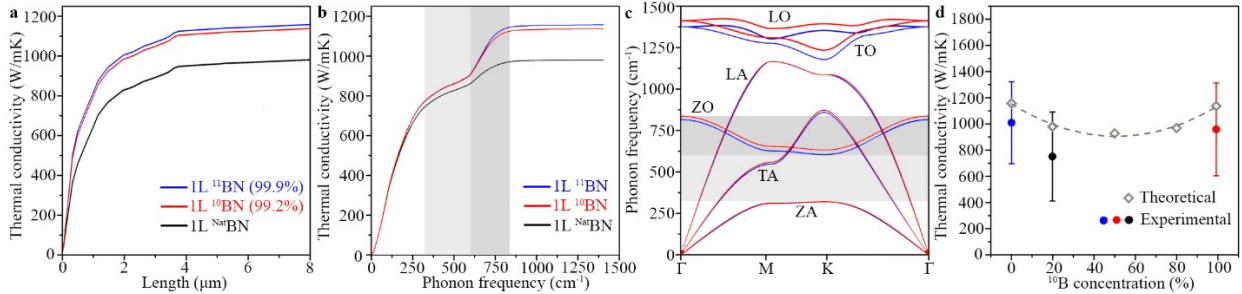

FIG. 3. (a) Theoretical cumulative $\kappa$ of 1L $^{10}$BN (99.2%), $^{11}$BN (99.9%), and $^{Nat}$BN as a function of sample length; (b) theoretical cumulative $\kappa$ of the same three materials as a function of phonon frequency, and the two phonon-frequency regions where $\kappa$ splits from different phonon contributions (see text for details) between 1L naturally occurring and monoisotopic BN are highlighted; (c) phonon dispersion of 1L $^{10}$BN (red) and $^{11}$BN (blue) with the same phonon-frequency regions of $\kappa$ divergence highlighted; (d) theoretical (open rhombus) and experimental (filled dots with standard deviations) $\kappa$ of 1L BN as a function of $^{10}$B concentration (i.e. 0.1%, 19.9%, 50%, 80%, and 99.2%), and corresponding parabolic fitting (dashed line).

Theoretical calculations were used to comprehensively understand the isotope effects. In the *ab initio* calculations of the $\kappa$ of 1L $^{10}$BN (99.2% $^{10}$B), $^{Nat}$BN, and $^{11}$BN (99.9% $^{11}$B), phonon-phonon, isotope, and boundary scatterings were taken into account. The boundary scattering rate was calculated as $v_g/L$, where $v_g$ is the group velocity of the phonons, and $L$ is the boundary length. Isotope mixing caused isotope scattering and shortened phonon mean free path ($\lambda$). In excellent agreement with our experimental results, the 99.9% $^{11}$BN had a slightly higher $\kappa$ than the 99.2% $^{10}$BN at 4 μm length close to the experimental sample size (FIG. 3a), revealing that the higher isotope purity in $^{11}$BN was the main cause of the slightly higher $\kappa$. FIG. 3b shows the theoretical accumulative $\kappa$ as a function of phonon frequency, and for comparison purpose, the phonon dispersions of 1L $^{10}$BN and $^{11}$BN are displayed in FIG. 3c. Apparently, the out-of-plane acoustic phonons (ZA) contributed to most of $\kappa$ in all BN sheets, consistent with our previous study [7]. Interestingly, 1L $^{10}$BN, $^{Nat}$BN, and $^{11}$BN showed almost no difference in the accumulative $\kappa$ at phonon frequency lower than 320 cm$^{-1}$, indicating that the isotope mixing had a small influence on the ZA phonons (FIG. 3c). In the region of 320−600 cm$^{-1}$ (highlighted in light grey in FIG. 3b-c), the accumulative $\kappa$ started to split between 1L naturally occurring and monoisotopic BN. This means that the isotopic purification decreased the scattering of transverse acoustic (TA) and longitudinal acoustic (LA) phonons. Nevertheless, the most significant deviation in $\kappa$ happened at ~600−800 cm$^{-1}$, corresponding well to the contribution of the out-of-plane optical phonons (ZO) (highlighted in dark grey in FIG. 3b-c). This suggests that the ZO phonon-isotope scatterings play an important role in the thermal conductivity of $^{Nat}$BN. Optical phonons barely contribute by themselves to thermal conductivity in bulk materials [45,46]; however as dimensionality is reduced, an important scattering channel for acoustic phonons is mediated through their optical counterparts. We observed that the majority of the difference in $\kappa$ between 1L naturally occurring and monoisotopic BN was caused by a strong acoustic-optical phonon scattering between ZO and TA/LA phonons due to isotope mixing [46]. FIG. 3d compares the $\kappa$ of 1L BN with different $^{10}$B concentrations, which followed a parabolic trend. For isotope scattering, $\lambda \propto g^{-1}T^{-4}$, where $g = \sum_i C_i \left[\left(M_i^2 - (\sum C_i M_i)\right)/(\sum C_i M_i)^2\right]^2$; $T$ is the temperature; $C_i$ and $M_i$ are the concentration and mass of isotope atoms, respectively. $g$ reached its maximum at about 50% $^{10}$B, resulting in the minimum $\lambda$ and hence $\kappa$. Our theoretical $\kappa$ values were larger than the experimental values and other theoretical predictions based on classical potentials (see Supporting Information, Table S1), and this could be due to the local density approximation (LDA). LDA is well known to over bind systems, leading to overestimations of phonon frequencies and consequently thermal conductivity [47].

We also measured the $\kappa$ of few-layer monoisotopic BN using the same procedure. They had smaller $\kappa$ values than those of the corresponding monolayers: 880±174 and 833±209 W/mK for 2L and 3L $^{10}$BN, and 930±204 and 902±193 W/mK for 2L and 3L $^{11}$BN, respectively, as shown in FIG. S7. The smaller $\kappa$ values were attributed to the increased Umklapp scattering of ZA phonons [7]. It is worth noting that the differences in $\kappa$ among 1-3L monoisotopic BN (*i.e.* $^{10}$BN: 1 : 0.92 : 0.87; $^{11}$BN: 1 : 0.92 : 0.89) were smaller than those among 1-3L $^{Nat}$BN (1 : 0.86 : 0.80) [7]. This can be explained by the larger contribution ratio of the ZA phonons to the overall thermal conductivity of $^{Nat}$BN than monoisotopic BN, because as aforementioned, reducing isotope mixing does not affect the ZA phonons much but significantly increases the contribution of TA/LA phonons to the thermal conductivity (FIG. 3b). The additional layers in 2-3L BN have a much larger influence on the ZA phonon scatterings than TA and LA phonons (FIG. S8). As a result, the increased Umklapp scatterings of ZA phonons due to the additional layers give rise to more prominent decreases in the thermal conductivity of few-layer $^{Nat}$BN than monoisotopic BN.

In summary, high-quality and suspended atomically thin isotopically pure BN sheets were produced by mechanical exfoliation, and their intrinsic in-plane thermal conductivities were measured by the optothermal Raman technique: 958±355 and 1009±313 W/mK for 1L $^{10}$BN and $^{11}$BN at close-to room temperature, respectively. These values were about 34% and 140% larger than those of 1L and bulk $^{Nat}$BN, respectively, attributed to 1) the longer mean free path of phonons mainly due to less ZO phonon-isotope scattering and subsequent reduced acoustic-optical scatterings between ZO and TA/LA phonons; 2) decreased phonon Umklapp scatterings in atomically thin samples caused by less interlayer interactions and hence reduced phonon branches. With its layered structure, low density, wide bandgap, excellent mechanical flexibility and strength, good chemical and thermal stability, atomically thin monoisotopic BN is promising on heat management in vdW devices and flexible electronics. This study not only deepens the fundamental understanding of isotope effect on 2D thermal conductivity but also forms the basis for further research and applications of isotopically engineered 2D materials.

ACKNOWLEDGMENT

L.H. Li thanks Prof. Takashi Taniguchi and Dr. Kenji Watanabe from National Institute for Materials Science, Japan for providing $^{Nat}h$BN single crystals for comparison; L.H.Li thanks the financial support from Australian Research Council (ARC) via Discovery Early Career Researcher Award (DE160100796). Q.Cai acknowledges ADPRF from Deakin University. E.J.G.Santos acknowledges the use of computational resources from the UK Materials and Molecular Modelling Hub for access to THOMAS supercluster, which is partially funded by EPSRC (EP/P020194/1) and the Cirrus UK National Tier-2 HPC Service at EPCC funded by the University of Edinburgh and EPSRC (EP/P020267/1) under contract ec019. The Queen's Fellow Award through the grant number M8407MPH, and the Department for the Economy (USI 097) are also acknowledged. Support from the Materials Engineering and Processing program of the National Science Foundation, award number CMMI 1538127, and the II−VI Foundation for monoisotopic *h*BN crystal growth is greatly appreciated. Part of the work was done at the Melbourne Centre for Nanofabrication (MCN) in the Victorian Node of the Australian National Fabrication Facility (ANFF) and on the soft X-ray beamline at the Australian Synchrotron, Victoria, Australia.